# Entangler via Electromagnetically Induced Transparency with Atomic Ensembles

Xihua Yang[1], Yuanyuan Zhou[1], and Min Xiao[2,3]

[1]*Department of Physics, Shanghai University, Shanghai 200444, China*

[2]*National Laboratory of Solid State Microstructures and Department of Physics, Nanjing University, Nanjing 210093, China*

[3]*Department of Physics, University of Arkansas, Fayetteville, Arkansas 72701, USA*

(Dated: Feb. 5, 2013)

We present an efficient and convenient scheme to entangle multiple optical fields via electromagnetically induced transparency in an atomic ensemble. The atomic spin wave, produced through electromagnetically induced transparency in the Λ-type configuration in an atomic ensemble, can be described by a Bose operator and acts as the entangler. By using the entangler, any desired number of nondegenerate narrow-band continuous-variable entangled fields, in principle, can be generated through stimulated Raman scattering processes, which holds great promise for applications in scalable quantum communication and quantum networks.

PACS numbers: 42.50.Gy, 03.67.Bg, 42.50.Dv, 42.65.Lm

Quantum state exchange between light and matter is a basic component for quantum interface in quantum information processing. As is well known, light is the best long-distance quantum information carrier and the atomic ensembles provide the promising tool for quantum information storage. In quantum information processing and quantum networks, generation of light-light, atom-atom, and atom-light multipartite entanglements plays an essential role in the implementation of quantum information protocols [1-3]. So far, the majority studies on entanglement have dealt with the generation of multiple entangled light fields. Apart from the conventional way of generating multipartite entanglement by mixing squeezed fields created through parametric down-conversion processes in nonlinear optical crystals with linear optical elements, i.e., polarizing beam splitters (PBS) [4, 5] as entanglers, the atomic ensembles provide an alternative avenue to the generation of multi-entangled fields due to the virtue of narrow bandwidth, nondegenerate frequencies, and long correlation time [6-9]. Based on the seminal proposal of Duan *et al.* [6], the electromagnetically induced transparency [10] (EIT)-based double-Λ-type atomic systems have been actively implemented for efficiently creating nondegenerate entangled twin fields through either nondegenerate four-wave mixing (FWM) or Raman scattering processes [7-8]. The multicolor multipartite continuous-variable (CV) entanglement has also been achieved by using the multiorder coherent Raman scattering [11] or multiple nondegenerate FWM processes [9]. Moreover, generating entanglement between the atomic ensembles and light fields, as well as between two atomic ensembles, have also been realized [6, 12-15], which are vital for storage and processing of quantum information.

In this paper, we propose an efficient and convenient scheme for quantum entangler via EIT in an atomic ensemble. The atomic spin wave, which can be described by a Bose operator and acts as the entangler, is produced through EIT in the Λ-type atomic configuration. Through stimulated Raman scattering processes, nondegenerate narrow-band multiple entangled fields upto any desired number can, in principle, be achieved by using such an entangler. The proposed entangler is quite distinct as compared to the conventional PBS entangler [16], since only coherent input light fields are needed for generating multipartite entanglement in the present scheme, whereas nonclassical input light fields are required by using the PBS entangler. Moreover, the present proposal is different from the previous ones as proposed in Refs. [9, 11]. Here, the entangled feature between the atomic ensemble and the generated fields is investigated and it is shown that under the EIT condition the generated atomic spin wave can serve as an entangler. Also, this proposal is superior to that proposed in Ref. [6], since, in principle, any desired number of entangled fields can be produced through this entangler.

The considered model is shown in Fig. 1a, where levels $|1\rangle$, $|2\rangle$, and $|3\rangle$, forming the $\Lambda$-type three-level system, correspond, respectively, to the ground-state hyperfine levels $5S_{1/2}$ (F=2), $5S_{1/2}$ (F=3), and the excited state $5P_{1/2}$ in $D_1$ line of $^{85}Rb$ atom with the ground-state hyperfine splitting of 3.036 GHz. The strong coupling field $E_c$ (with frequency $\omega_c$ and Rabi frequency $\Omega_c$) and relatively weak probe field $E_p$ (with frequency $\omega_p$ and Rabi frequency $\Omega_p$) are tuned to resonance with the transitions $|2\rangle$-$|3\rangle$ and $|1\rangle$-$|3\rangle$, respectively. By applying a third mixing field $E_{m1}$ (with frequency $\omega_{m1}$ and Rabi frequency $\Omega_{m1}$), off-resonantly coupling levels $|1\rangle$ and $|3\rangle$, a Stokes field $E_1$ can be created through nondegenerate FWM process [9]. In fact, the produced Stokes field can be equivalently regarded as the result of scattering the mixing field $E_{m1}$ off the atomic spin wave (S) pre-established by the coupling and probe fields in the $\Lambda$-type EIT configuration formed by levels $|1\rangle$, $|2\rangle$, and $|3\rangle$, as shown in Fig. 1b, where the induced atomic spin wave acts as a frequency converter with frequency equal to the separation between the two lower states [17, 18]. Subsequently, if more laser fields $E_{m2}$, $E_{m3}$,….. $E_{mN}$ (with N being a positive integer) are applied to off-resonantly couple levels $|1\rangle$ and $|3\rangle$, more Stokes fields $E_2$, $E_3$,….. $E_N$ can be produced through scattering the mixing fields off the same atomic spin wave. In what follows, we will investigate the entanglement feature between the atomic spin wave and the generated Stokes fields.

In the EIT-based configuration, we assume that the Rabi frequency of the scattering field is far smaller than its frequency detuning, and the atomic spin wave is strong enough to ensure that different scattering fields have little influence on it. We also assume that the probe field is relatively weak as compared to the strong coupling field. Under these conditions, the collective atomic spin field S ($S = (1/\sqrt{N_a})\sum_i |1\rangle_{i\ i}\langle 2|$ with $N_a$ being the total number of atoms in the interaction volume) can be viewed as approximately satisfying the bosonic commutation relation $[S, S^+] = (\sum_i |1\rangle_{i\ i}\langle 1| - \sum_i |2\rangle_{i\ i}\langle 2|)/\sqrt{N_a} \approx 1$, so the produced atomic spin wave S can be treated as a bosonic field and the quantum state of the atomic ensemble can be transferred to that of the generated Stokes fields [18].

We first investigate the bipartite entanglement between the atomic spin wave S and the generated Stokes field $E_1$. As shown in Fig. 1b, we assume that the generated Stokes field is very weak as compared to the scattering field, thus, the scattering field is treated classically, whereas the Stokes field $E_1$ and atomic spin field S are treated quantum mechanically. After adiabatic elimination of the upper excited state, the effective Hamiltonian of the system in the interaction picture can be written as [18]

$$H_I = i\hbar(k_1 a_1^+ S^+ - k_1 a_1 S),$$

where $a_1$ is the bosonic annihilation operator for the Stokes field $E_1$, and $k_1 = g_{23}\Omega_{m1}\sqrt{N_a}/\Delta_1$ with $\Delta_1 = \omega_{m1} - \omega_{31}$ as the detuning of the mixing field $E_{m1}$ from the resonant transition $|1\rangle - |3\rangle$, and $g_{23}$ being the coupling coefficient between the Stokes field and its respective atomic states. According to the previous experimental parameters [9], $k_1$ is in the order of several cm$^{-1}$; in such case, the estimated timescale for $k_1 t = 1$ corresponds to about tens of ps. By solving the Heisenberg equations of motion for the operators $S(t)$ and $a_1(t)$, one can get the solutions for the two operators as functions of their initial values:

$$S(t) = \cosh(k_1\sqrt{a}t)S(0) + \sqrt{a}\sinh(k_1\sqrt{a}t)a_1^+(0),$$

$$a_1(t) = \frac{1}{\sqrt{a}}\sinh(k_1\sqrt{a}t)S^+(0) + \cosh(k_1\sqrt{a}t)a_1(0),$$

where $a = \cos^2\theta - \sin^2\theta$ with $tg\theta = \Omega_p/\Omega_c$. As seen in Fig. 1b, the collective atomic state is initially in a coherent superposition state, and the Stokes field is initially in vacuum, so the initial state of the atom-field system can be written as $|\varphi_0\rangle = (1/\sqrt{N_a})\sum_i(\cos\theta|1\rangle_i + \sin\theta|2\rangle_i)|0_s\rangle$, where $|1\rangle|0_s\rangle$ ($|2\rangle|0_s\rangle$) represents an atom in state $|1\rangle$ ($|2\rangle$) and the Stokes field in vacuum. We use the criterion $V = (\Delta u)^2 + (\Delta v)^2 < 4$ proposed by Duan *et al.* [19] to verify the bipartite entanglement between the atomic spin wave S and the generated Stokes field $E_1$, where $u = x_1 - x_S$ and $v = p_1 + p_S$ with $x_j = (a_j + a_j^+)$ and $p_j = -i(a_j - a_j^+)$. Smaller the correlation V is, stronger the degree of the bipartite entanglement becomes.

Figure 2 displays the interaction time evolutions of V under different ratios of $\Omega_p/\Omega_c$ with $k_1=1$. Obviously, it can be seen in Figs. 2a-2d that bipartite entanglement can be created as soon as the interaction is nonzero; under different ratios of $\Omega_p/\Omega_c$, V, with the initial value of about 4, evolves with interaction time and becomes less than 4, which sufficiently indicates the generation of genuine bipartite entanglement between the Stokes field $E_1$ and the atomic spin field S. In addition, there exists a range of interaction time that V nearly approaches to zero, which means that the Stokes field is perfectly quantum correlated with the atomic spin wave. With the increase of the ratio $\Omega_p/\Omega_c$ in a moderate range, the interaction time range of near-zero V values would decrease and the degree of bipartite entanglement would be weakened.

The stimulated Raman scattering process described above for generating bipartite entanglement between the Stokes field $E_1$ and the atomic spin field S has the similar feature as the parametric down-conversion process in a nonlinear optical crystal, where the generated Stokes field $E_1$ can be equivalently viewed as the result of frequency down-conversion process through mixing the scattering field $E_{m1}$ with the atomic spin wave S pre-established by the coupling and probe fields. As the generation of a Stokes photon is always accompanied by an atomic spin-wave excitation, the down-converted frequency component (i.e., Stokes field) is strongly quantum correlated with the atomic spin wave; as a result, strong bipartite entanglement between the Stokes field and the atomic spin wave can be established. This idea can be generalized to the case with more laser fields $E_{m2}$, $E_{m3}$,..... $E_{mN}$ tuned to the vicinity of the transition $|1\rangle$-$|3\rangle$ to mix with the same atomic spin wave, then all of the generated Stokes fields $E_1$, $E_2$,..... $E_N$ would be entangled with the atomic spin wave, and therefore entangled with each other. In such situation any desired number of entangled light fields, in principle, can be created with the atomic spin wave acting as an entangler. This proposed scheme can be demonstrated by applying another laser field $E_{m2}$ scattering off the atomic spin wave to generate tripartite entanglement.

In the following, we consider the case of generating the Stokes fields $E_1$ and $E_2$ through scattering the laser fields $E_{m1}$ and $E_{m2}$ off the atomic spin wave, respectively, as shown in Fig. 1b. The effective Hamiltonian of the system in the interaction picture has the form [18]

$$H_I = i\hbar \left[ k_1 \left( a_1^+ S^+ - a_1 S \right) + k_2 \left( a_2^+ S^+ - a_2 S \right) \right],$$

where $k_2 = g_{23}\Omega_{m2}\sqrt{N_a}/\Delta_2$ with $\Delta_2 = \omega_{m2} - \omega_{31}$ being the frequency detuning of the mixing field $E_{m2}$ from the transition $|1\rangle$-$|3\rangle$. In a similar way as calculating the bipartite entanglement generation, by solving the Heisenberg equations of motion for the operators $S(t)$, $a_1(t)$, and $a_2(t)$, the solutions for the three operators as functions of their initial values can be expressed as:

$$S(t) = \cosh(\beta t)S(0) + \frac{k_1 a}{\beta}\sinh(\beta t)a_1^+(0) + \frac{k_2 a}{\beta}\sinh(\beta t)a_2^+(0),$$

$$a_1(t) = \frac{k_1}{\beta}\sinh(\beta t)S^+(0) + [\frac{k_1^2 a}{\beta^2}(\cosh\beta t - 1) + 1]a_1(0) + [\frac{k_1 k_2 a}{\beta^2}(\cosh\beta t - 1)]a_2(0),$$

$$a_2(t) = \frac{k_2}{\beta}\sinh(\beta t)S^+(0) + [\frac{k_1 k_2 a}{\beta^2}(\cosh\beta t - 1)]a_1(0) + [\frac{k_2^2 a}{\beta^2}(\cosh\beta t - 1) + 1]a_2(0),$$

where $\beta = \sqrt{(k_1^2 + k_2^2)a}$. We demonstrate the tripartite entanglement of the generated Stokes fields $E_1$, $E_2$, and atomic spin wave S by using the criterion proposed by van Lock-Furusawa (VLF) [5] with inequalities as follows:

$$V_{12} = V(x_1 - x_2) + V(p_1 + p_2 + g_S p_S) < 4,$$

$$V_{1S} = V(x_1 - x_S) + V(p_1 + g_2 p_2 + p_S) < 4,$$

$$V_{2S} = V(x_2 - x_S) + V(g_1 p_1 + p_2 + p_S) < 4,$$

where $V(A) = \langle A^2 \rangle - \langle A \rangle^2$ and $g_i$ (i=1, 2, 3) are arbitrary real numbers. In a similar way to that used in Ref. [20], we set $g_1 = \frac{-(\langle p_1 p_2 \rangle + \langle p_1 p_S \rangle)}{\langle p_1^2 \rangle}$, $g_2 = \frac{-(\langle p_1 p_2 \rangle + \langle p_2 p_S \rangle)}{\langle p_2^2 \rangle}$, and $g_S = \frac{-(\langle p_1 p_S \rangle + \langle p_2 p_S \rangle)}{\langle p_S^2 \rangle}$. Satisfying any pair of these three inequalities is a sufficient condition for realizing tripartite entanglement, and also smaller the correlations $V_{12}$, $V_{1s}$, and $V_{2s}$ are, higher the degree of the tripartite entanglement will be.

Figures 3a-3d depict the evolutions of the VLF correlations as a function of interaction time with $k_1 = 1$ and $\Omega_p / \Omega_c = 1/20$ under different $k_2$ values. It can be seen that when $k_2$ is far smaller than $k_1$ (see Fig. 3a), there only exists a very limited range of the interaction time within which the inequalities for $V_{12}$ and $V_{2s}$ are fulfilled, whereas $V_{1s}$ with the minimal value nearly equal to zero is smaller than 4 in a wide range of the interaction time; this means that strong entanglement between the generated Stokes field $E_1$ and atomic spin wave is obtained, whereas the generated Stokes field $E_2$ is weakly entangled with both the Stokes field $E_1$ and atomic spin wave S. With the increase of $k_2$, the minimal values of $V_{12}$ and $V_{2s}$ would decrease and the range of interaction time with both $V_{12}$ and $V_{2s}$ being smaller than 4 would increase, whereas $V_{1s}$ displays an opposite behavior, that is, the degree of bipartite entanglement between the generated Stokes field $E_1$ and atomic spin wave is weakened and that between the Stokes field $E_2$ and Stokes field $E_1$ (or atomic spin wave) is strengthened. When $k_1=k_2=1$ (see Fig. 3c), $V_{12}$, approaching to 2 after certain interaction time, is smaller than 4 over the whole interaction time, and there exists a broad range of interaction time within which the inequalities for $V_{12}$, $V_{1s}$, and $V_{2s}$ are all satisfied. In this case, the Stokes fields $E_1$ and $E_2$ and the atomic spin wave S are CV entangled with each other, and the strongest tripartite entanglement is obtained. Further increasing $k_2$ would enhance the degree of bipartite entanglement between the Stokes field $E_2$ and atomic spin wave, and degrade the entanglement between the Stokes field $E_1$ and Stokes field $E_2$ (or atomic spin wave). In principle, this scheme can be generalized to generate any desired number of maximally-entangled fields by using the atomic spin wave as an entangler through applying more laser fields $E_{m3}$, $E_{m4}$,….. $E_{mN}$ tuned to the vicinity of the transition $|1\rangle$-$|3\rangle$ under the condition of $k_1=k_2=k_3….=k_N$. Detailed calculations show that the VLF correlations $V_{12}$, $V_{1s}$, and $V_{2s}$ exhibit weak dependence on the variation of ratio $\Omega_p/\Omega_c$, as long as the probe field is relatively weak as compared to the coupling field.

In the above EIT-based configuration, the Rabi frequencies of the scattering fields should be small as compared to their frequency detunings so that the coupling between different scattering fields can be neglected, and the atomic spin wave should be strong enough to ensure that different scattering fields have little influence on it, which can be realized by using substantially strong probe and coupling fields as compared to the scattering fields. Moreover, the probe field should be relatively weak as compared to the coupling field, so that the produced atomic spin wave S can be treated as a bosonic field. In an atomic medium, there will be realistic imperfections, such as finite coherence times, atomic diffusion, and Doppler broadening. In the above analysis, we did not take into account these effects, so that the main physics can be resolved. The influences to entanglement due to those realistic imperfections will be analyzed in the future work by using a Heisenberg-Langevin approach.

.It should be noted that the present scheme is different from that proposed by Duan *et al*. [6]. In the current scheme, an atomic spin wave is created in advance via

EIT, which acts as an entangler and provides a way to generate nondegenerate multiple CV entangled fields to any desired number. In addition, in Ref. [6], the atomic spin wave is produced through spontaneous Raman scattering, so the generated Stokes or anti-Stokes field would emit in all directions and the photon production efficiency is very low, whereas in the present scheme, the generated Stokes field would propagate along a particular direction determined by the phase-matching condition for FWM [9], and the photon production efficiency is much higher, which would bring great facility in realistic quantum information processing protocals.

Note also that, as compared to the routinely-employed method to produce multi-field entanglement by using PBS entangler [4-5], where the interferometric stabilization of the optical paths is required, and the entangled multi-fields are degenerate and suffer from short correlation time (~ps), the present scheme can be employed to generate narrow-band multicolor entangled CV fields with long correlation time (~ms or even ~s [21]) by using a single entangler, where the correlation time is determined by the coherence decay time between the two lower states due to the finite interaction time between atoms and light. Furthermore, the proposed entangler is different from and superior to the PBS entangler [16] in that the generation of multipartite entanglement using PBS entangler is a linear process, whereas it is a nonlinear one using the present entangler, so as nonclassical input light fields are required for using PBS entangler to generate multipartite entanglement, only coherent input light fields are needed for using the present entangler to create multipartite entanglement, which will greatly simplify the practical implementation.

In conclusion, we have proposed an efficient and convenient scheme for entangler via electromagnetically induced transparency with an atomic ensemble. The entangler, in principle, has the virtue of being able to generate arbitrary number of nondegenerate and narrow-band CV-entangled fields with long correlation time, which are the essential elements for a scalable quantum repeater for the realization of long-distance quantum communication. In addition, only coherent input light fields are required for generating multipartite entanglement. Together with the entanglement generations between two atomic ensembles and entanglement swapping [6, 12-15], light-light, atom-atom, and atom-light multipartite entanglements can be achieved, which will find promising applications in quantum information processing and quantum networks.

## ACKNOWLEDGEMENTS

This work is supported by NBRPC (Nos. 2012CB921804 and 2011CBA00205), the National Natural Science Foundation of China (Nos. 11274225, 10974132, 50932003, and 11021403), and Innovation Program of Shanghai Municipal Education Commission (No. 10YZ10). Yang's e-mail is yangxih@yahoo.com.cn; M. Xiao's e-mail is mxiao@uark.edu.

## REFERENCES

[1] D. Bouweester, A. Ekert, and A. Zeilinger, *The Physics of Quantum Information* (Springer-Verlag, Berlin, 2000).

[2] M.A. Nielsen, and I.L. Chuang, *Quantum Computation and Quantum Information* (Cambridge University Press, Cambridge, 2000).

[3] H.J. Kimble, Nature **453**, 1023 (2008).

[4] E. Knill, R. Laflamme, and G.J. Milburn, Nature **409**, 46 (2001); J. Jing, *et al.*, Phys. Rev. Lett. **90**, 167903 (2003); T. Aoki, *et al.*, Phys. Rev. Lett. **91**, 080404 (2003).

[5] P. van Loock and S.L. Braunstein, Phys. Rev. Lett. **84**, 3482 (2000); P. van Loock and A. Furusawa, Phys. Rev. A **67**, 052315 (2003).

[6] L.M. Duan, M.D. Lukin, J.I. Cirac, and P. Zoller, Nature **414**, 413 (2001).

[7] C. H. van der Wal, *et al.*, Science **301**, 196 (2003); A. Kuzmich, *et al.*, Nature **423**, 731 (2003).

[8] V. Balic, *et al.*, Phys. Rev. Lett. **94**, 183601 (2005); V. Boyer, *et al.*, Science **321**, 544 (2008); H. Wu, and M. Xiao, Phys. Rev. A **80**, 063415 (2009).

[9] X.H. Yang, J.T. Sheng, U. Khadka, M. Xiao, Phys. Rev. A **85**, 013824 (2012); X.H. Yang, Y.Y. Zhou, M. Xiao, Phys. Rev. A **85**, 052307 (2012).

[10] E. Arimondo, Prog. Opt. **35**, 257 (1996); S.E. Harris, Phys. Today **50** (9), 36 (1997); M. Xiao, *et al.*, Phys. Rev. Lett. **74**, 666 (1995); M. Fleischhauer, A. Imamoglu, J.P. Marangos, Rev. Mod. Phys. **77**, 633 (2005).

[11] F.L. Kien, A.K. Patnaik, and K. Hakuta, Phys. Rev. A **68**, 063803 (2003).

[12] L.M. Duan, *et al.*, Phys. Rev. Lett. **85**, 5643 (2000).

[13] B. Julsgaard, A. Kozhekin, and E.S. Polzik, Nature **413**, 400 (2001).

[14] Z.S. Yuan, *et al.*, Nature **454**, 1098 (2008).

[15] N. Sangouard, *et al.*, Rev. Mod. Phys. **83**, 33 (2011), and references therein.

[16] M.S. Kim, *et al.*, Phys. Rev. A 65, 032323 (2002); X.B. Wang, Phys. Rev. A **66**, 024303 (2002); **66**, 064304 (2002); X.L. Feng, R.X. Li, and Z.Z. Xu, Phys. Rev. A **71**, 032335 (2005).

[17] A.J. Merriam, *et al.*, Phys. Rev. Lett. **84**, 5308 (2000).

[18] Z.Y. Ou, Phys. Rev. A **78**, 023819 (2008).

[19] L.M. Duan, *et al.*, Phys. Rev. Lett. **84**, 2722 (2000).

[20] M.K. Olsen and A.S. Bradley, Phys. Rev. A **74**, 063809 (2006).

[21] R. Zhang, S. Garner, and L.V. Hau, Phys. Rev. Lett. **103**, 233602 (2009); U. Schnorrberger, *et al.*, Phys. Rev. Lett. **103**, 033003 (2009).

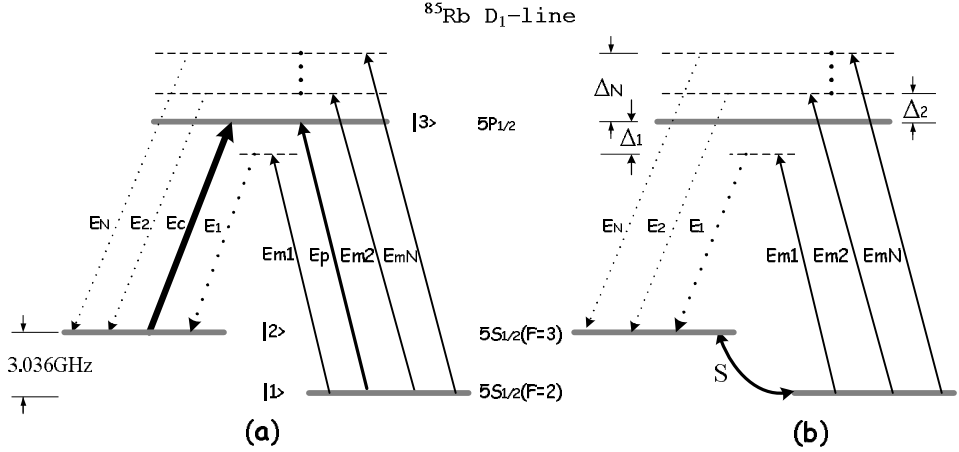

FIG. 1. (**a**) The Λ-type system of the $D_1$ transitions in $^{85}$Rb atom coupled by the coupling ($E_c$), probe ($E_p$), and mixing ($E_{m1}$, $E_{m2}$,….. $E_{mN}$) fields and the corresponding Stokes fields ($E_1$, $E_2$,….. $E_N$) generated through multiple FWM processes. (**b**) The equivalent configuration of (**a**) with the two lower states driven by the atomic spin wave S prebuilt by the strong on-resonant $E_c$ and $E_p$ fields in the Λ-type EIT configuration.

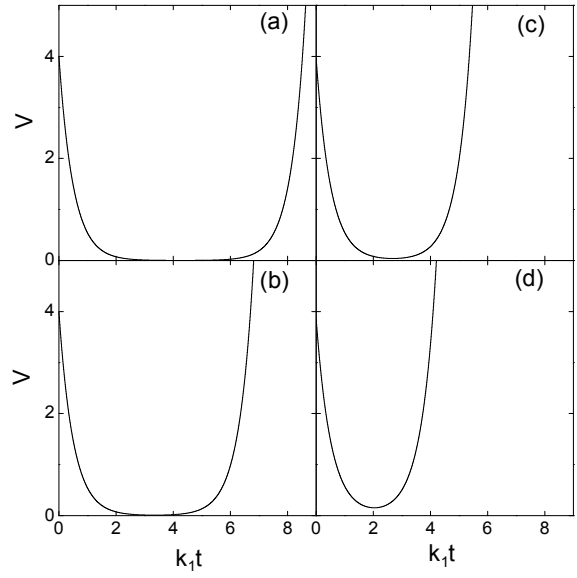

FIG. 2. The evolutions of V as a function of interaction time in terms of the normalized time $k_1 t$ under different ratio $\Omega_p / \Omega_c$ with $k_1 = 1$. **(a)** $\Omega_p / \Omega_c = 1/50$, **(b)** $\Omega_p / \Omega_c = 1/20$, **(c)** $\Omega_p / \Omega_c = 1/10$, and **(d)** $\Omega_p / \Omega_c = 1/5$.

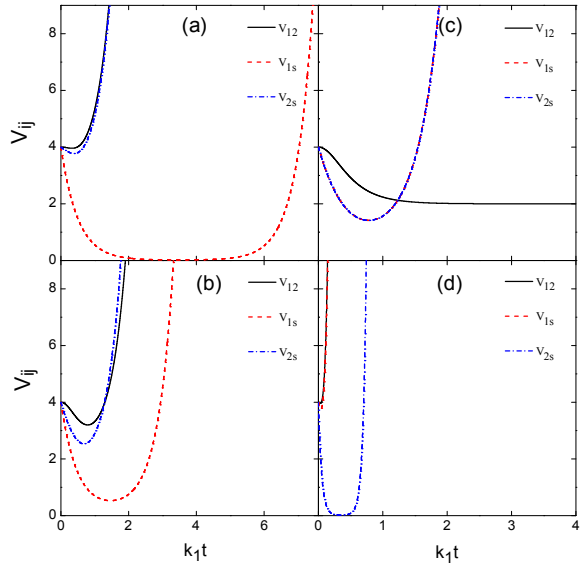

FIG. 3. The evolutions of the VLF correlations $V_{12}$, $V_{1S}$, and $V_{2S}$ as a function of interaction time under different $k_2$ with $k_1=1$ and $\Omega_p/\Omega_c = 1/20$. (**a**) $k_2=0.1$, (**b**) $k_2=0.5$, (**c**) $k_2=1$, and (**d**) $k_2=10$.